\documentclass[12pt,draftclsnofoot,onecolumn]{IEEEtranTCOM}
\IEEEoverridecommandlockouts

\ifCLASSOPTIONcompsoc
  \usepackage[nocompress]{cite}
\else
  \usepackage{cite}
\fi

\ifCLASSINFOpdf
\else
\fi

\makeatletter
\renewcommand\subsubsection{\@startsection{subsubsection}{3}{\z@}%
                       {-18\p@ \@plus -4\p@ \@minus -4\p@}%
                       {4\p@ \@plus 2\p@ \@minus 2\p@}%
                       {\normalfont\normalsize\bfseries\boldmath
                        \rightskip=\z@ \@plus 8em\pretolerance=10000 }}
\renewcommand\paragraph{\@startsection{paragraph}{4}{\z@}%
                       {-12\p@ \@plus -4\p@ \@minus -4\p@}%
                       {2\p@ \@plus 1\p@ \@minus 1\p@}%
                       {\normalfont\normalsize\itshape
                        \rightskip=\z@ \@plus 8em\pretolerance=10000 }}
\makeatother

\usepackage{xcolor}
\usepackage{soul}
\usepackage{graphicx}
\usepackage{hyperref}
\usepackage{bm}
\usepackage{amsmath,amssymb}
\usepackage{indentfirst}
\let\labelindent\relax
\usepackage{enumitem}
\usepackage{multirow}
\usepackage{subcaption}
\DeclareMathOperator*{\argmax}{argmax}

\usepackage{algorithm}
\usepackage{amsthm}
\usepackage{tabularx}
\usepackage{array,booktabs}
\usepackage{dutchcal}
\newcolumntype{C}{>{$\displaystyle}c<{$}}
\usepackage[noend]{algpseudocode}

\newcolumntype{P}[1]{>{\centering\arraybackslash}p{#1}}
\graphicspath{{figures/}}
\makeatletter
\newcommand{\multiline}[1]{%
  \begin{tabularx}{\dimexpr\linewidth-\ALG@thistlm}[t]{@{}X@{}}
    #1
  \end{tabularx}
}
\makeatother
\begin{document}
\title{Learning to Estimate: A Real-Time Online Learning Framework for MIMO-OFDM Channel Estimation}
\author{Lianjun Li, Sai Sree Rayala, Jiarui Xu, Lizhong Zheng, and Lingjia Liu
\thanks{L. Li, S. Rayala, J. Xu, and L. Liu are with Wireless@Virginia Tech, Bradley Dept. of Electrical and Computer Engineering at Virginia Tech. L. Zheng is with the EECS Department at Massachusetts Institute of Technology. 
This work was supported in part by US National Science Foundation (NSF) under grants: CCF-1937487, CNS-2003059, and CNS-2002908.
The corresponding author is L. Liu (ljliu@ieee.org).}
}

\maketitle
\begin{abstract}
In this paper we introduce StructNet-CE, a novel real-time online learning framework for MIMO-OFDM channel estimation, which only utilizes over-the-air (OTA) pilot symbols for online training and converges within one OFDM subframe.
The design of StructNet-CE leverages the structure information in the MIMO-OFDM system, including the repetitive structure of modulation constellation and the invariant property of symbol classification to inter-stream interference.
The embedded structure information enables StructNet-CE to conduct channel estimation with a binary classification task and accurately learn channel coefficients with as few as two pilot OFDM symbols. Experiments show that the channel estimation performance is significantly improved with the incorporation of structure knowledge. StructNet-CE is compatible and readily applicable to current and future wireless networks, demonstrating the effectiveness and importance of combining machine learning techniques with domain knowledge for wireless communication systems.

\end{abstract}


\section{Introduction}\label{sec:intro}

Multiple-input multiple-output with orthogonal frequency division multiplexing (MIMO-OFDM) has been adopted as one of the core technologies in 4G LTE and 5G NR to meet the ever-increasing traffic volume demand in modern wireless communication. 
The core benefits MIMO-OFDM brings to modern wireless system are twofold: 1) By utilizing spatial diversity and multiplexing techniques, MIMO can improve transmission reliability and system throughput; 2) OFDM divides frequency band into multiple subcarriers to convert a broad-band frequency-selective fading channel into narrow-band flat fading channels, thus enabling simple and efficient transmission.
One key step for realizing the full potential of MIMO-OFDM is the acquisition of accurate channel state information (CSI), with which various precoding, beamforming, and symbol detection techniques can be deployed to utilize the desired features as well as compensate the undesired effects in wireless systems. 

In commercial wireless systems such as cellular and WiFi networks, transmitters send known pilot/reference signals to receivers for channel estimation purpose. Regarding pilot-based channel estimation, least square (LS) and linear minimum mean square error (LMMSE) are two popular methods. LS \cite{kay1993fundamentals, van1995channel,hou2005unified} channel estimation is conducted through a matrix pseudo-inverse, it doesn't require any channel statistics or noise variance as prior information, thus widely adopted in practical systems due to its simplicity. However, it has low estimation accuracy due to the well-know noise amplification effect. On the other hand, if channel statistics and noise variance are known, LMMSE methods \cite{van1995channel,edfors1998ofdm,miao2004space,zhang2005channel, luo2008general} can provide better channel estimates by utilizing channel correlation information. It is worth to mention that MIMO-OFDM channel has a three-dimensional structure lies in space, time, and frequency domain, so LMMSE methods can utilize correlation among all three dimensions or a subset of them. For example, \cite{zhang2005channel} considers the space-frequency correlation, while \cite{luo2008general} formulates correlation across all three dimensions. Although LMMSE is optimal in terms of minimizing the mean square error (MSE), the channel statistics are not easy to acquire in practical wireless systems, which limits its application scenario.

Motivated by the limitations of conventional methods and recent success of machine learning, researchers are looking into learning-based approaches for channel estimation, with the hope that neural networks (NNs) can perform better than LS, while not requiring the knowledge of channel statistics. For example, A 3-hidden-layer multilayer perceptron (MLP) is  designed in \cite{ye2017power} for jointly channel estimation and demodulation.
A convolutional neural network (CNN) is adopted in~\cite{neumann2018learning} to learn the parameters of the minimum mean square error (MMSE) channel estimator.
Two deep neural networks (DNNs) are designed in~\cite{huang2018deep} to facilitate the direction-of-arrival (DOA) estimation and channel gain estimation for the massive MIMO system.
\cite{he2018deep} proposed a denoising CNN-based method for beamspace channel estimation in millimeter wave massive MIMO system.
A DNN-based algorithm is introduced in~\cite{gizzini2020deep} to improve the coarse channel estimation of spectral temporal averaging method in IEEE 802.11p system. 
Recent works also view channel estimation as an image super-resolution problem, where LS channel estimate is treated as low-resolution image and fed into NN, the output of NN is refined channel estimate, i.e., the high-resolution image.
Specifically, ChannelNet \cite{soltani2019deep} combines an image super-resolution network with an image denoising network to improve the LS channel estimation quality.
\cite{balevi2019deep} modifies the denoising network proposed in \cite{heckel2018deep} to perform channel estimation.
ReEsNet \cite{li2019deep} further improves the channel estimation performance by adopting residual learning-based design for the NN. Due to the promising performance demonstrated by attention mechanism \cite{vaswani2017attention} and transformer \cite{devlin2018bert} in nature language processing, researchers also investigate attention-based designs for channel estimation task \cite{chen2020channel, mashhadi2021pruning, gao2021attention}. 

Although aforementioned learning-based algorithms do not require channel statistics as prior knowledge, they all require offline training, meaning the NNs are trained offline by artificially generated data, once training is complete, the learnt NN weights stay fixed during the online inference phase. The offline training methodology only works under the assumption that the offline training data has the same statistical information as the online testing one, when they are statistically different, the channel estimation performance cannot be guaranteed, which is known as ``uncertainty in generalization''~\cite{Shafin2020}. This is the main drawback that prevents offline learning-based channel estimation methods from being adopted in any practical system. 

Given the limitation of offline learning-based solutions, a natural research choice would be developing purely online learning-based algorithms. However, unlike offline training where the noise-free ground truth channel is available in the offline data and utilized as training label, in online scenario there is no way to obtain the true channel. Therefore, how to obtain online training label becomes the key problem of designing purely online method. Some works try to tackle this issue by designing new training loss \cite{zheng2021online,jha2021online}, for example, inspired by the conventional LS channel estimation loss (which doesn't depend on the ground truth channel, the loss is the received pilot signal minus transmitted pilot signal times estimated channel), \cite{zheng2021online} introduces an online training loss that maps the LS loss onto a lower dimension space that leverages the rank-restricted isometry property of the massive MIMO channel. Other work try to avoid using online training label by adopting reinforcement learning (RL), a conventional successive denoising algorithm is proposed in \cite{oh2021channel} with RL agent to learn the denoising sequence, although this method can be classified as online method, it requires channel power and number of channel taps as prior knowledge. Moreover, all aforementioned online methods require hundreds of OFDM subframes for the training to converge.

In this work we introduce StructNet-CE, a novel learning framework for channel estimation task, which is 1) online, only OTA pilot symbols are needed for the NN training; and 2) real-time, the training converges within one OFDM subframe. To be specific, 
\begin{itemize}
    \item By explicitly incorporating wireless channel coefficients into the symbol detection problem formulation and corresponding NN design, our StructNet-CE can learn the channel estimation task through symbol detection loss.
    \item Unlike other works that treat NN as a black-box, our NN design leverages wireless communication domain knowledge, namely the QAM symbol repetitive pattern and the interference invariant property, making StructNet-CE a compact NN model with fast convergence. 
\end{itemize}

The design of StructNet-CE naturally introduces a new methodology for learning-based channel estimation, which completely differentiates it from all prior works.
The real-time online learning capability makes it a relevant and essential strategy for practical MIMO-OFDM systems, including 5G and beyond networks.

The rest of the paper is organized as follows. Section \ref{sec:system} Introduces MIMO-OFDM transceiver procedure and conventional channel estimation methods. Section \ref{sec:structnet} explains the design of StructNet-CE, as well as its training procedure. Experiment results are discussed in section \ref{sec:experiment}. Section \ref{sec:conclusion} concludes this paper. 

\textbf{Notations}: $\mathbb{C}(\mathbb{R})$ denotes the complex (real) number set. $\mathbb{Z}$ denotes the integer set. Scalar, vector, and matrix are denoted by non-bold letter, bold lowercase letter, and bold uppercase letter respectively, e.g., $x$, $\bm{x}$, and $\bm{X}$. $(\cdot)^{'}$, and $(\cdot)^*$ denotes respectively the transpose, and Hermitian transpose operation. $\hat{\bm{A}}$ is the estimation of matrix $\bm{A}$. $[\bm{A}_1 \ \bm{A}_2]$ denotes concatenate matrices $\bm{A}_1$ and $\bm{A}_2$ along the column dimension, while $[\bm{A}_1; \bm{A}_2]$ or $\begin{bmatrix}\bm{A}_1\\ \bm{A}_2\end{bmatrix}$ denotes concatenate those two matrices along the row dimension.

\section{System Model} \label{sec:system}
In this section, we first introduce MIMO-OFDM transceiver procedure, then talk about conventional channel estimation methods. Table. \ref{tab:notation} summarizes MIMO-OFDM related notations.

\subsection{MIMO-OFDM Transceiver Procedure}
\begin{figure}[htb]
    \centering
    \includegraphics[width=0.65\columnwidth]{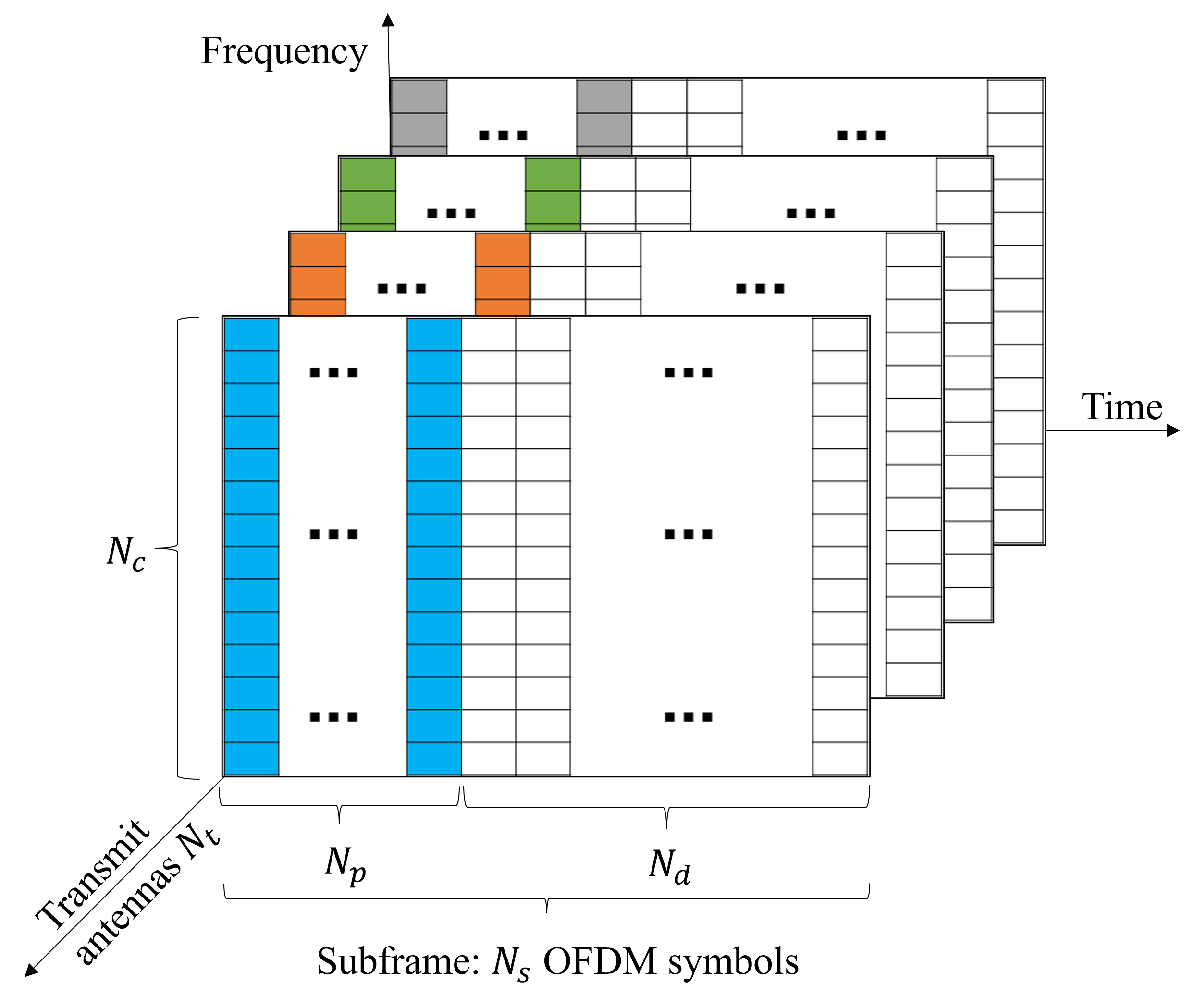}
    \caption{MIMO-OFDM subframe structure}
    \label{fig:frame}
\end{figure}

We consider a MIMO-OFDM system with $N_t$ transmit antennas (data streams) and $N_r$ receive antennas. As depicted in Fig. \ref{fig:frame}, the OFDM subframe consists of $N_c$ subcarriers and $N_s$ OFDM symbols, within which the first $N_p$ OFDM symbols are pilot symbols designed for channel estimation purpose, and the rest $N_d = N_s-N_p$ OFDM symbols are data symbols. 
The OFDM symbols in frequency-domain are first converted to time-domain by an inverse fast Fourier transform (IFFT), then a cyclic prefix (CP) with length $N_{\mathrm{cp}}$ is inserted to avoid the inter-symbol interference (ISI) caused by the multi-path wireless channel.
At receiver side, the time-domain signal is converted to frequency-domain by first removing CP and then performing a fast Fourier transform (FFT). The relationship between transmitted and received symbols at subcarrier $c \in \{0, \dots, N_c-1\}$ can be expressed as
\begin{equation}
    \bm{Y}(c) = \bm{H}(c)\bm{X}(c) +\bm{N}(c),
    \label{eq:MIMO}
\end{equation}
where $\bm{X}(c) \in \mathbb{C}^{N_t \times N_s}$ is the transmitted quadrature amplitude modulation (QAM) symbol at subcarrier $c$; $\bm{Y}(c) \in \mathbb{C}^{N_r \times N_s}$ is the corresponding received signal; $\bm{H}(c)\in \mathbb{C}^{N_r \times N_t}$ represents the frequency domain channel at subcarrier $c$; $\bm{N}(c)$ is the additive white Gaussian noise (AWGN). In this paper we consider the block-fading scenario where the channel remains constant over $N_s$ OFDM symbols within a subframe and varies across subframes.

To recover transmitted symbols, the receiver first performs channel estimation by utilizing $N_p$ pilot symbols, several conventional channel estimation methods will be introduced later in \ref{sec:conventional-ce-method}. With estimated channel $\hat{\bm{H}}(c)$, then the transmitted symbols can be recovered by LMMSE equalization as:
\begin{equation}
    \hat{\bm{X}}(c) = \big(\hat{\bm{H}}^*(c)\hat{\bm{H}}(c) + \sigma^2\bm{I}_{N_t} \big)^{-1}\hat{\bm{H}}^*(c)\bm{Y}(c),\label{eq:symbol-detection}
\end{equation}
where $\sigma^2$ is the noise variance and $\bm{I}_{N_t} \in \mathbb{C}^{N_t \times N_t}$ is the identity matrix. 

\subsection{Conventional Channel Estimation Methods}\label{sec:conventional-ce-method}
\paragraph{Least square (LS)}
Denote $\bm{X}_p(c) \in \mathbb{C}^{N_t \times N_p}$, and $\bm{Y}_p(c) \in \mathbb{C}^{N_r \times N_p}$ as transmitted and received pilot symbols at subcarrier $c$, LS channel estimation can be obtained as \cite{kay1993fundamentals}:
\begin{equation}
    \hat{\bm{H}}_{LS}(c) = \bm{Y}_p(c) \bm{X}_p^*(c) \big(\bm{X}_p(c)  \bm{X}_p^*(c) \big)^{-1}.\label{eq:LS}
\end{equation}
LS is an easy-to-implement method which doesn't require any prior knowledge on the channel statistics and noise variance, it can only provide mediocre accuracy due to the well-known noise amplification effect.

\paragraph{Linear minimum mean square error (LMMSE)}
If the second order channel statistics and noise variance are known, a better channel estimation can be obtained by LMMSE method \cite{Ozdemir2007channel}, which in essence is a filtering operation on LS channel estimation. Denote $\hat{\bm{h}}_{LS}^{r,t} \in \mathbb{C}^{N_c}$ as the frequency-domain LS channel estimation between receiving antenna $r$ and transmitting antenna $t$ across all $N_c$ subcarriers, the LMMSE channel estimation $\hat{\bm{h}}_{LMMSE}^{r,t} \in \mathbb{C}^{N_c}$ can be calculated as:
\begin{equation}
    \hat{\bm{h}}_{LMMSE}^{r,t} = \bm{R}_{hh}^{r,t}(\bm{R}_{hh}^{r,t} + \sigma^2\bm{I}_{N_c})^{-1}\hat{\bm{h}}_{LS}^{r,t}, \label{eq:LMMSE}
\end{equation}
where $\bm{R}_{hh}^{r,t} \triangleq \mathbb{E}[\bm{h}^{r,t}(\bm{h}^{r,t})^*]\in \mathbb{C}^{N_c \times N_c}$ is the correlation matrix of channel $\bm{h}^{r,t} \in \mathbb{C}^{N_c}$, and $\bm{I}_{N_c} \in \mathbb{C}^{N_c \times N_c}$ is the identity matrix.

\paragraph{Empirical LMMSE (em-LMMSE)}
It can be seen from equation (\ref{eq:LMMSE}) that LMMSE requires exact channel correlation statistics as prior knowledge, which is difficult to obtain in a practical system. A viable solution is to use empirical statistics instead of the exact one, we name this method em-LMMSE and it is expressed as:
\begin{equation}
    \hat{\bm{h}}_{em-LMMSE}^{r,t} = \hat{\bm{R}}_{hh}^{r,t}(\hat{\bm{R}}_{hh}^{r,t} + \sigma^2\bm{I}_{N_c})^{-1}\hat{\bm{h}}_{LS}^{r,t}, \label{eq:em-LMMSE}
\end{equation}
where $\hat{\bm{R}}_{hh}^{r,t}$ is the empirical channel correlation matrix, it is initialized as an identity matrix and updated based on moving average of previously estimated channel correlations, i.e., 
\begin{equation}
    \hat{\bm{R}}_{hh}^{r,t} \triangleq \mathbb{E}\big[\hat{\bm{h}}^{r,t}_{em-LMMSE}(\hat{\bm{h}}^{r,t}_{em-LMMSE})^*\big]
\end{equation}

\begin{table}
\centering
\caption{MIMO-OFDM related notations}
\label{tab:notation}
\resizebox{1\columnwidth}{!}{%
\begin{tabular}{|c|c|c|}
\hline
\textbf{Symbols} &\textbf{Data type \& shape} & \textbf{Definitions}  \\ \hline
$N_r$ & $\mathbb{R}^{1}$ & Number of receiver antennas  \\ \hline
$N_t$ & $\mathbb{R}^{1}$ & Number of transmitter antennas\\ \hline
$N_{c}$ & $\mathbb{R}^{1}$ & Number of OFDM subcarriers\\ \hline
$N_{cp}$ & $\mathbb{R}^{1}$ & Length of Cyclic Prefix (CP)\\ \hline
$N_{p}$ & $\mathbb{R}^{1}$ & Number of pilot symbols in one OFDM subframe\\ \hline
$N_{d}$ & $\mathbb{R}^{1}$ & Number of data symbols in one OFDM subframe\\ \hline
$N_{s}$ & $\mathbb{R}^{1}$ & $N_{p} + N_{d}$\\ \hline
$\bm{X}(c)$ & $\mathbb{C}^{N_t \times N_s}$ & Transmitted symbols at subcarrier $c$\\ \hline
$\bm{Y}(c)$ & $\mathbb{C}^{N_r \times N_s}$ & Received symbols at subcarrier $c$\\ \hline
$\bm{H}(c)$ & $\mathbb{C}^{N_r \times N_t}$ & Wireless channel at subcarrier $c$\\ \hline
$\bm{X}_p(c)$ & $\mathbb{C}^{N_t \times N_p}$ & Transmitted pilot symbols at subcarrier $c$\\ \hline
$\bm{Y}_p(c)$ & $\mathbb{C}^{N_r \times N_p}$ & Received pilot symbols at subcarrier $c$\\ \hline
$\hat{\bm{H}}_{LS}(c)$ & $\mathbb{C}^{N_r \times N_t}$ & LS estimated channel at subcarrier $c$\\ \hline
$\bm{h}^{r,t}$ & $\mathbb{C}^{N_c}$ & Wireless channel between transmit antenna $t$ and receive antenna $r$ across all subcarriers\\ \hline
$\hat{\bm{h}}_{LMMSE}^{r,t}$ & $\mathbb{C}^{N_c}$ & LMMSE estimation of channel $\bm{h}^{r,t}$\\ \hline
$\bm{R}_{hh}^{r,t}$ & $\mathbb{C}^{N_c \times N_c}$ & Correlation matrix of channel  $\bm{h}^{r,t}$\\ \hline
$\hat{\bm{h}}_{em-LMMSE}^{r,t}$ & $\mathbb{C}^{N_c}$ & Empirical LMMSE estimation of channel $\bm{h}^{r,t}$\\ \hline
$\hat{\bm{R}}_{hh}^{r,t}$ & $\mathbb{C}^{N_c \times N_c}$ & Empirical correlation matrix of channel $\bm{h}^{r,t}$\\ \hline
\end{tabular}%
}
\end{table}

\section{StructNet for Channel Estimation} \label{sec:structnet}
StructNet was initially introduced in \cite{xu2022rc} for MIMO symbol detection task. Later in \cite{Xu2022Struct} a channel layer was added to the NN for better symbol detection performance. Note this channel layer contains wireless channel coefficients that can be updated by the symbol detection loss during training, this observation motivates us to utilize StructNet and further extend it with interference invariant property to construct a new learning framework for channel estimation, which is 1) online, the NN is trained by OTA pilot symbols instead of offline data required by most of learning-based channel estimation methods; and 2) real-time, this method converges within one subframe, in contrast to other online learning methods require large number of consecutive subframes to converge. We name this channel estimation framework StructNet-CE.

In the rest of this section, we first explain our method from theoretic point of view, then introduce the neural network design, finally discuss the training procedure. 

\subsection{Theoretic explanation}

MIMO symbol detection can be treated as a multinomial classification problem, by explicitly incorporating wireless channel coefficients into the problem formulation, StructNet-CE can perform channel estimation through a symbol detection task. Specifically, wireless channels are utilized to form two special properties of the symbol detection problem: 1) \textit{shifting property}, through shifting the received signal along the desired channel direction to certain positions determined by the QAM constellation symbol interval, the multinomial classification problem can be solved with a single binary classifier; 2) \textit{interference invariant property}, shift the received signal along the interference channel direction should not affect the symbol detection on the desired stream.

Let's consider the MIMO signal on one subcarrier and one OFDM symbol, for notation simplification we drop the subcarrier index $c$ for now, the MIMO model in equation (\ref{eq:MIMO}) can be rewritten as
\begin{equation*}
    \bm{y} = \bm{H}\bm{x} + \bm{n},
\end{equation*}
where $\bm{x} \in \mathcal{A}^{N_t}$ are the transmitted symbols, $\mathcal{A}$ is the QAM constellation set, e.g., for 16-QAM, $\mathcal{A} = \{-3,-1,+1,+3\}\times \{-3j,-1j,+1j,+3j\}$. $\bm{H} \in \mathbb{C}^{N_r \times N_t}$ represents the wireless channel. $\bm{y} \in \mathbb{C}^{N_r}$ are the received symbols. $\bm{n}$ is noise. Denote $x_i$ as the $i$th element of $\bm{x}$, and $\bm{h}_i$ as the $i$th column of $\bm{H}$, when performing symbol detection on data stream $x_i$, we call $\bm{h}_i$ the desired channel, and all other channels $\bm{h}_j (j\neq i)$ the interference channels.    

The real-valued version of $\bm{x}$ and $\bm{y}$ are used for the symbol detection task, which are defined as:
\begin{equation*}
    \bm{\tilde{x}} \triangleq 
    \begin{bmatrix}
    \text{Re}(\bm{x})\\
    \text{Im}(\bm{x})
    \end{bmatrix},\ \ \text{and} \ \  
    \bm{\tilde{y}} \triangleq 
    \begin{bmatrix}
    \text{Re}(\bm{y})\\
    \text{Im}(\bm{y})
    \end{bmatrix},
\end{equation*}
where $\bm{\tilde{x}}$ and $\bm{\tilde{y}}$ now represent transmitted and received 4-PAM symbols.
The symbol detection can be expressed as a posteriori estimation problem:
\begin{equation}
    \argmax_{\bm{\tilde{x}}}P(\bm{\tilde{x}}|\bm{\tilde{y}}),
\end{equation}
denote the $i$th element of $\bm{\tilde{x}}$ as $\tilde{x}_i$, by applying naive Bayesian principle, the joint distribution $P(\bm{\tilde{x}}|\bm{\tilde{y}})$ can be approximated with marginal distribution $P_i(\tilde{x}_i|\bm{\tilde{y}})$:
\begin{equation}
    P(\bm{\tilde{x}}|\bm{\tilde{y}}) \approx \prod_{i=1}^{2N_t}P_i(\tilde{x}_i|\bm{\tilde{y}}),
\end{equation}
then the symbol detection can be done by maximizing marginal distributions:
\begin{equation}
     \argmax_{\tilde{x}_i}P_i(\tilde{x}_i|\bm{\tilde{y}}), \ 1\le i\le 2N_t.
\end{equation}
To solve this problem, we design and train NNs to approximate $P_i(\tilde{x}_i|\bm{\tilde{y}})$, i.e., after training,
\begin{equation}
    f_i(\tilde{x}_i;\bm{\tilde{y}}) \approx P_i(\tilde{x}_i|\bm{\tilde{y}})
\end{equation}
where $f_i(\tilde{x}_i;\bm{\tilde{y}})$ denotes the NN with input $\bm{\tilde{y}}$ and output corresponding to $\tilde{x}_i$.

\begin{figure}[htb]
    \centering
    \includegraphics[width=0.6\columnwidth]{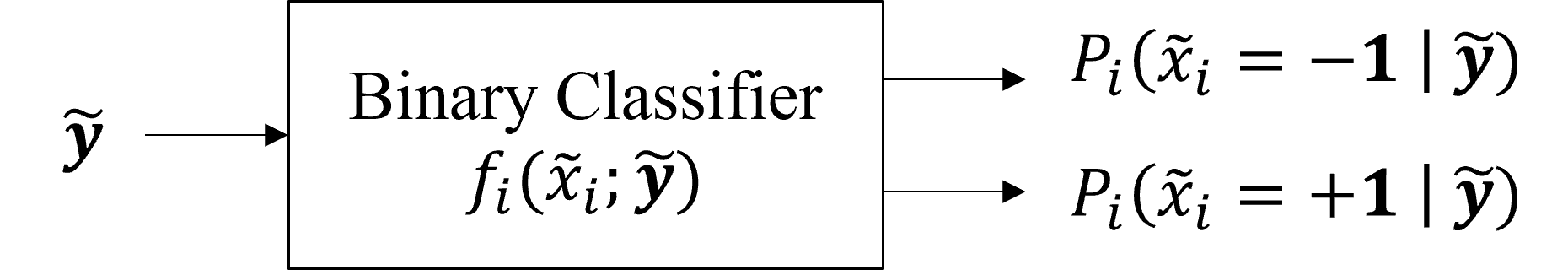}
    \caption{Binary classification}
    \label{fig:binary}
\end{figure}

\textbf{Binary classification:} let's consider a binary decision case first, where $\tilde{x}_i \in \{-1,+1\}$, then the NN is a binary classifier with two outputs, which are trained to estimate the probability of the corresponding two classes,
\begin{align*}
    f_i(\tilde{x}_i = -1;\bm{\tilde{y}}) &\approx P_i(\tilde{x}_i = -1|\bm{\tilde{y}}),\\
    f_i(\tilde{x}_i = +1;\bm{\tilde{y}}) &\approx P_i(\tilde{x}_i = +1|\bm{\tilde{y}}),
\end{align*}
when testing, the decision is made by choosing the class with higher probability. The NN process is depict in Fig. \ref{fig:binary}.

\begin{figure}[htb]
    \centering
    \includegraphics[width=0.65\columnwidth]{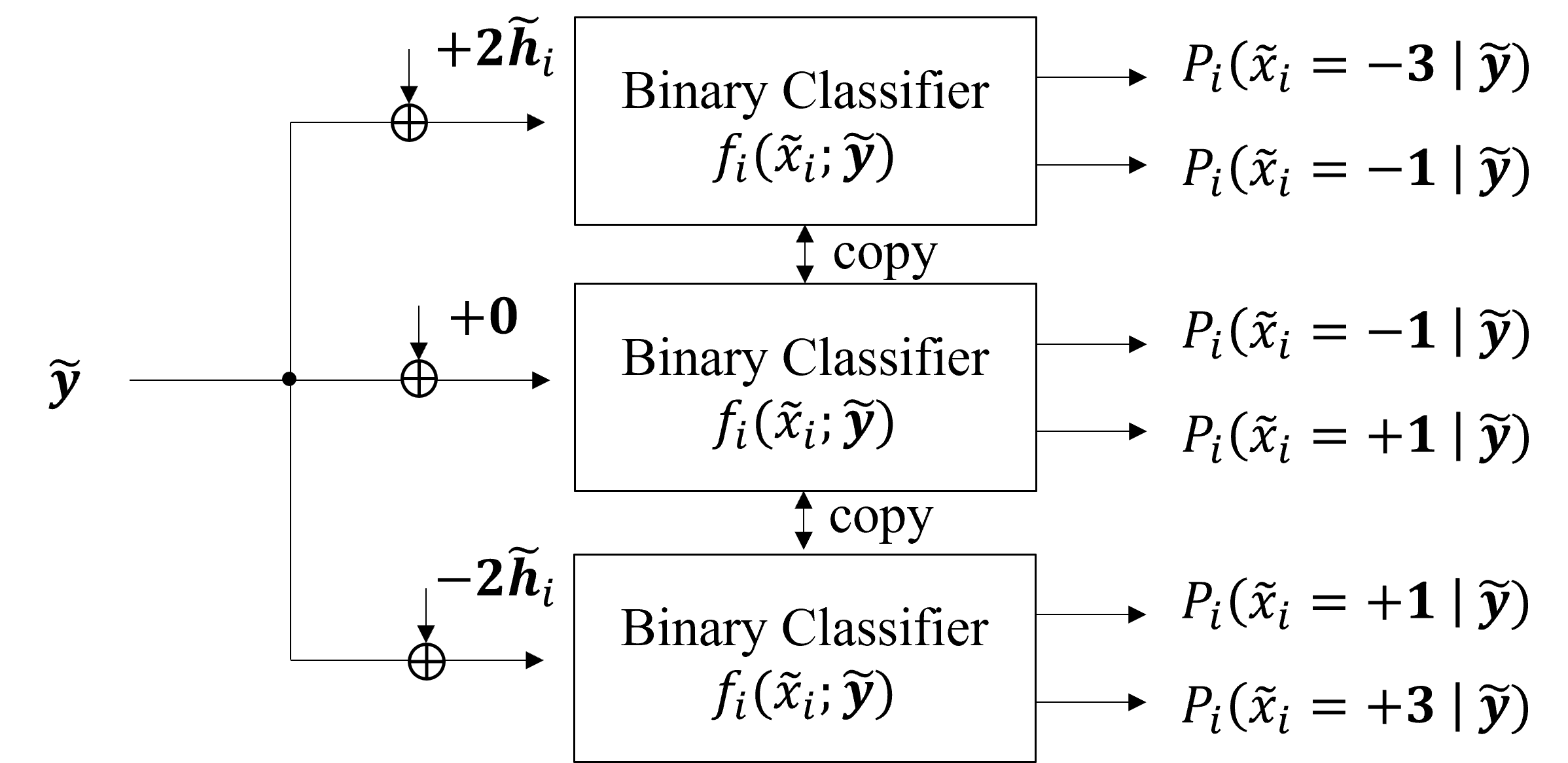}
    \caption{Multinominal classification}
    \label{fig:Multinominal-classification}
\end{figure}

\textbf{Multinominal classification:} when the transmitted symbol is not binary, e.g, $\tilde{x}_i \in \{-3,-1,+1,+3\}$, we can still utilize a single binary classifier to estimate the probabilities of all classes through the following shifting principle:
\begin{align}
    \frac{P_i(\tilde{x}_i = -3|\bm{\tilde{y}})}{P_i(\tilde{x}_i = -1|\bm{\tilde{y}})} &= \frac{P_i(\tilde{x}_i = -1|\bm{\tilde{y}}+2\bm{\tilde{h}}_i)}{P_i(\tilde{x}_i = +1|\bm{\tilde{y}}+2\bm{\tilde{h}}_i)}\\
    \frac{P_i(\tilde{x}_i = +1|\bm{\tilde{y}})}{P_i(\tilde{x}_i = +3|\bm{\tilde{y}})} &= \frac{P_i(\tilde{x}_i = -1|\bm{\tilde{y}}-2\bm{\tilde{h}}_i)}{P_i(\tilde{x}_i = +1|\bm{\tilde{y}}-2\bm{\tilde{h}}_i)},
\end{align}
where $+2\bm{\tilde{h}}_i$ and $-2\bm{\tilde{h}}_i$ are the shifting vectors. Let's take $+2\bm{\tilde{h}}_i$ as an example. It consists of three parts, a $+$ sign, meaning we are shifting a transmitted symbol to its right neighbor in the constellation, e.g., from $-3$ to $-1$; a scalar $2$, which is distance between those two symbols; and the desired channel vector $\bm{\tilde{h}}_i$ is defined as
\begin{equation}
    \bm{\tilde{h}}_i \triangleq \begin{cases}
    \big[\text{Re}(\bm{h_i}); \ \text{Im}(\bm{h_i})\big], \ &\text{if} \ 1\le i\le N_t\\
    \big[-\text{Im}(\bm{h_{i-N_t}}); \ \text{Re}(\bm{h_{i-N_t}})\big]; \ &\text{if} \ N_t < i \le 2N_t
    \end{cases}\label{eq:h_tilde}
\end{equation}
we name the sign and scalar together as shifting parameter, and use $\lambda_i$ to denote it when needed.
In summary, shift the received symbol $\bm{\tilde{y}}$ by $+2\bm{\tilde{h}}_i$ is equivalent to shift the transmitted symbol from -3 to -1, in this way the binary classifier can estimate the probability of class -3. Similarly, shift $\bm{\tilde{y}}$ by $-2\bm{\tilde{h}}_i$, the probability of class +3 can be estimated. With the binary classifier and shifted inputs, the probabilities of all classes can be obtained by solving below equations:
\begin{align}
    \frac{P_i(\tilde{x}_i = -3|\bm{\tilde{y}})}{P_i(\tilde{x}_i = -1|\bm{\tilde{y}})} &= \frac{f_i(\tilde{x}_i = -1;\bm{\tilde{y}}+2\bm{\tilde{h}}_i)}{f_i(\tilde{x}_i = +1;\bm{\tilde{y}}+2\bm{\tilde{h}}_i)},\nonumber\\
    \frac{P_i(\tilde{x}_i = -1|\bm{\tilde{y}})}{P_i(\tilde{x}_i = +1|\bm{\tilde{y}})} &= \frac{f_i(\tilde{x}_i = -1;\bm{\tilde{y}}+0)}{f_i(\tilde{x}_i = +1;\bm{\tilde{y}}+0)},\nonumber\\
    \frac{P_i(\tilde{x}_i = +1|\bm{\tilde{y}})}{P_i(\tilde{x}_i = +3|\bm{\tilde{y}})} &= \frac{f_i(\tilde{x}_i = -1;\bm{\tilde{y}}-2\bm{\tilde{h}}_i)}{f_i(\tilde{x}_i = +1;\bm{\tilde{y}}-2\bm{\tilde{h}}_i)},\nonumber\\
    \sum_{a=\{-3,-1,1,3\}} & P_i(\tilde{x}_i = a|\bm{\tilde{y}}) = 1. \label{eq:rcstruct-test}
\end{align}
The decision is made by choosing the class with the highest probability. The multinominal classification process is depict in Fig. \ref{fig:Multinominal-classification}, note the three binary classifiers are actually one, they are copies of each other.

\begin{figure}[htb]
    \centering
    \includegraphics[width=0.65\columnwidth]{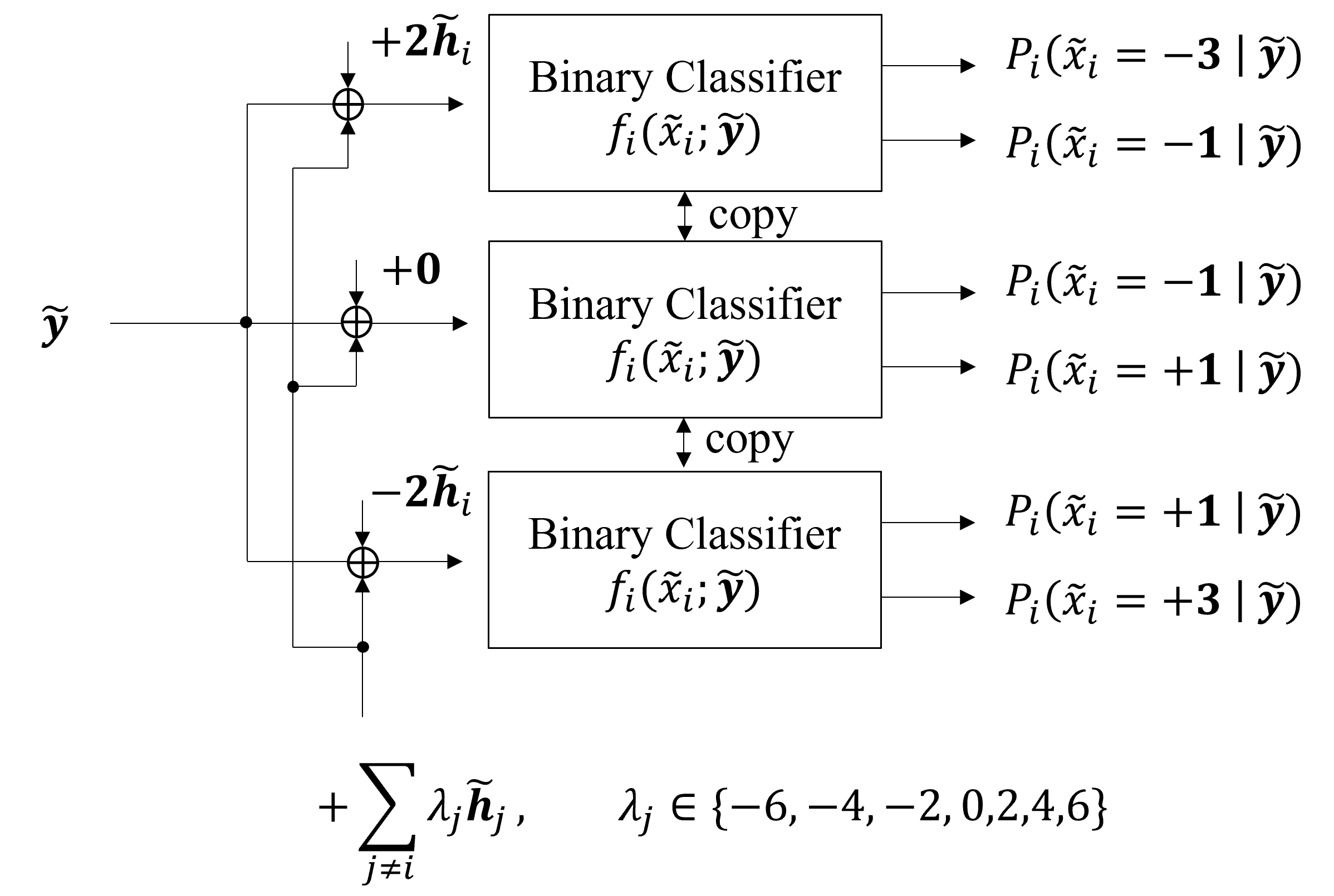}
    \caption{Multinominal classification with interference invariant property}
    \label{fig:Multinominal-classification-II}
\end{figure}

\textbf{Interference invariant}: when detecting the desired stream $\tilde{x}_i$, the result should be invariant to the transmit symbol changes in the interference streams. Again, take the 4-PAM case as an example, apply the shifting property on interference streams, we have:
\begin{equation}
    P_i\Big(\tilde{x}_i|\bm{\tilde{y}} + \sum_{j\neq i}\lambda_j\tilde{\bm{h}}_j\Big) = P_i(\tilde{x}_i|\bm{\tilde{y}}), \ \   \lambda_j \in \{-6, -4, -2, 0, 2, 4, 6\},
\end{equation}
and the trained NN should also has this property, i.e.,
\begin{equation}
        f_i\Big(\tilde{x}_i;\bm{\tilde{y}} + \sum_{j\neq i}\lambda_j\tilde{\bm{h}}_j\Big) = f_i(\tilde{x}_i;\bm{\tilde{y}}), \ \   \lambda_j \in \{-6, -4, -2, 0, 2, 4, 6\}.
\end{equation}
The multinominal classification process with interference invariant is illustrated in Fig. \ref{fig:Multinominal-classification-II}. There are two ways to realize this interference invariant property, one is augmenting the training data by shifting along the interference channel directions, then let the NN learn this property through training; Another is manually design a NN layer which is interference invariant, so the NN doesn't need to spend effort on learning what is already known --- the domain knowledge. We prefer the second option, more details will be discussed in the following section \ref{sec:NN-design}.

\subsection{Network Architecture} \label{sec:NN-design}
\begin{figure}[htb]
    \centering
    \includegraphics[width=0.8\columnwidth]{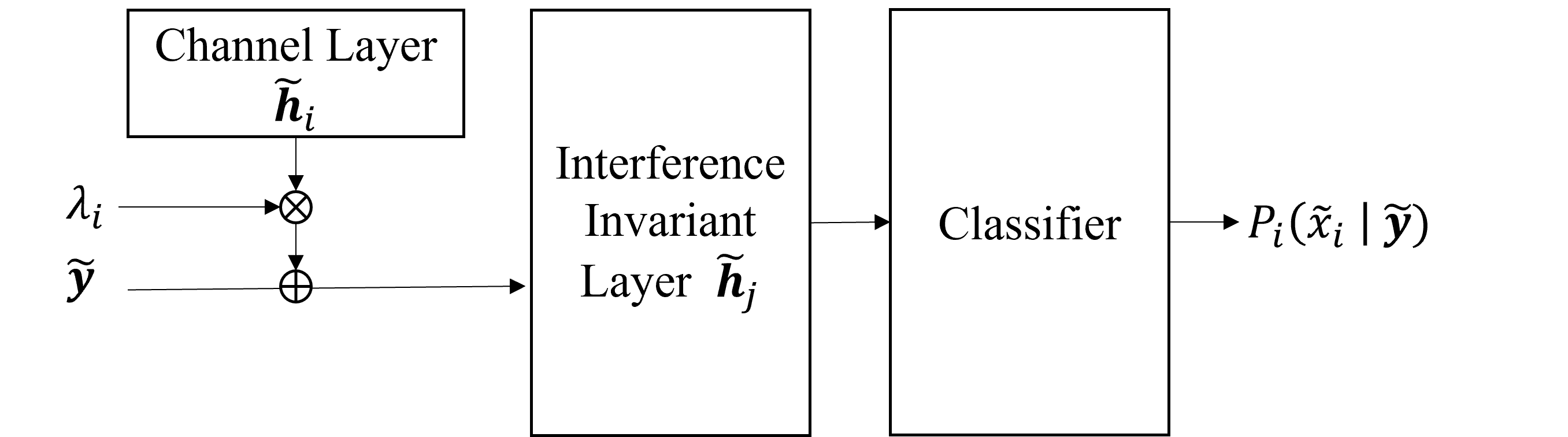}
    \caption{StructNet-CE}
    \label{fig:structnet}
\end{figure}

The NN architecture of StructNet-CE is illustrated in Fig. \ref{fig:structnet}. Compared with our previous work StructNet \cite{Xu2022Struct}, this work is a more advanced design that incorporates the interference invariant property. The inputs of NN are received signal $\tilde{\bm{y}}$, and shifting parameter $\lambda_i$. The desired channel coefficients $\tilde{\bm{h}}_i$ are stored in the channel layer. The shifted signal $\tilde{\bm{y}} + \lambda_i \tilde{\bm{h}}_i$ goes into the interference invariant layer (IIL). Then the output of IIL goes into the binary classifier and finally produces the NN output $P_i(\tilde{x}_i| \tilde{\bm{y}})$. The desired channel coefficients $\tilde{\bm{h}}_i$ and interference channel coefficients $\tilde{\bm{h}}_j$ are initialized by LS channel estimation, after being updated by the pilot symbol detection training loss, they are read out from corresponding layers and viewed as the NN estimated channel.

\paragraph{Channel layer}
The channel layer is implemented as a NN linear layer, with desired channel coefficients stored as its weights, which has size of 2$N_t$. The NN weights can be updated through back propagation of the cross-entropy loss at the output of the binary classifier.

\paragraph{Interference invariant layer}
As mentioned earlier, there are two options to realize the interference invariant property, one is augment the training data and let NN learn, another is manually design a IIL. The first option is simple and straightforward from design perspective, but it requires NN to spend computation resources on learning something we already know, which is not efficient from learning perspective. Therefore, we are more interested in the second option, where we design a NN layer which is naturally interference invariant. 

Denote $f_i^{\text{IIL}}(\bm{z})$ as the IIL with input $\bm{z}$, then the interference invariant property requires:
\begin{equation}
    f_i^{\text{IIL}} \Big(\bm{z} + \sum_{j\neq i} \lambda_j \tilde{\bm{h}}_j \Big) = f_i^{\text{IIL}} (\bm{z}), \ \ \lambda_j \in \{\cdots, -4, -2, 0, 2, 4, \cdots\}.
\end{equation}
From above equation we can see $f_i^{\text{IIL}}(\cdot)$ is actually a periodic function with period $2\tilde{\bm{h}}_j$, now the task becomes how to construct this periodic function, next we introduce two solutions.

\textbf{The shifting solution}: recall basic algebra knowledge that a periodic function can be constructed as summation of shifted versions of an arbitrary function, i.e., given an arbitrary function $g(\bm{z})$, a periodic function $f(\bm{z})$ with period $N$ can be constructed as:
\begin{equation}
    f(\bm{z}) = \sum_{m \in \mathbb{Z}} g(\bm{z} + mN).\label{eq:shift-period}
\end{equation}

\begin{figure}[tb]
    \centering
    \includegraphics[width=0.5\columnwidth]{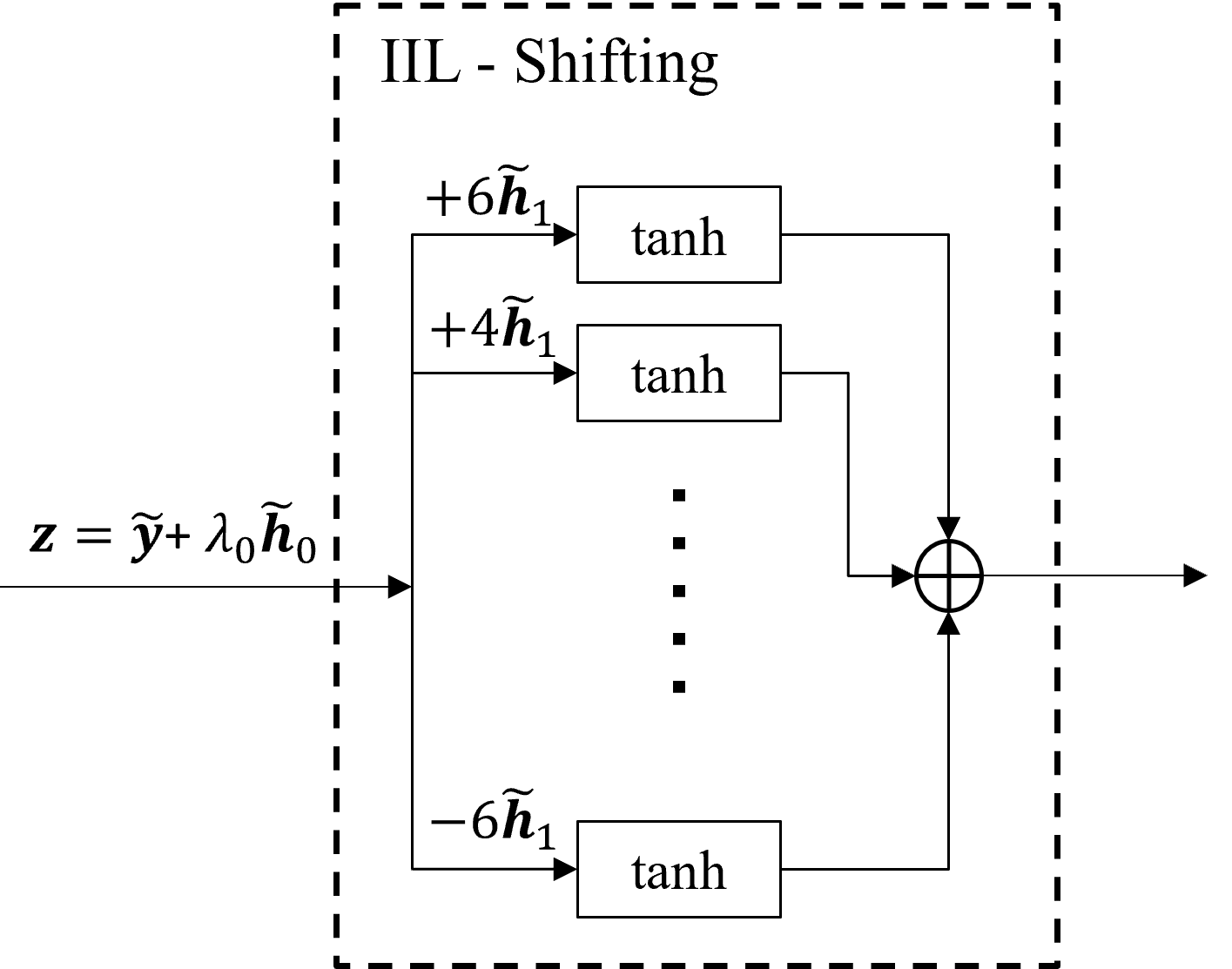}
    \caption{IIL by the shifting solution, two data streams}
    \label{fig:IIL-shift-2}
\end{figure}

\begin{figure}[htb]
    \centering
    \includegraphics[width=0.45\columnwidth]{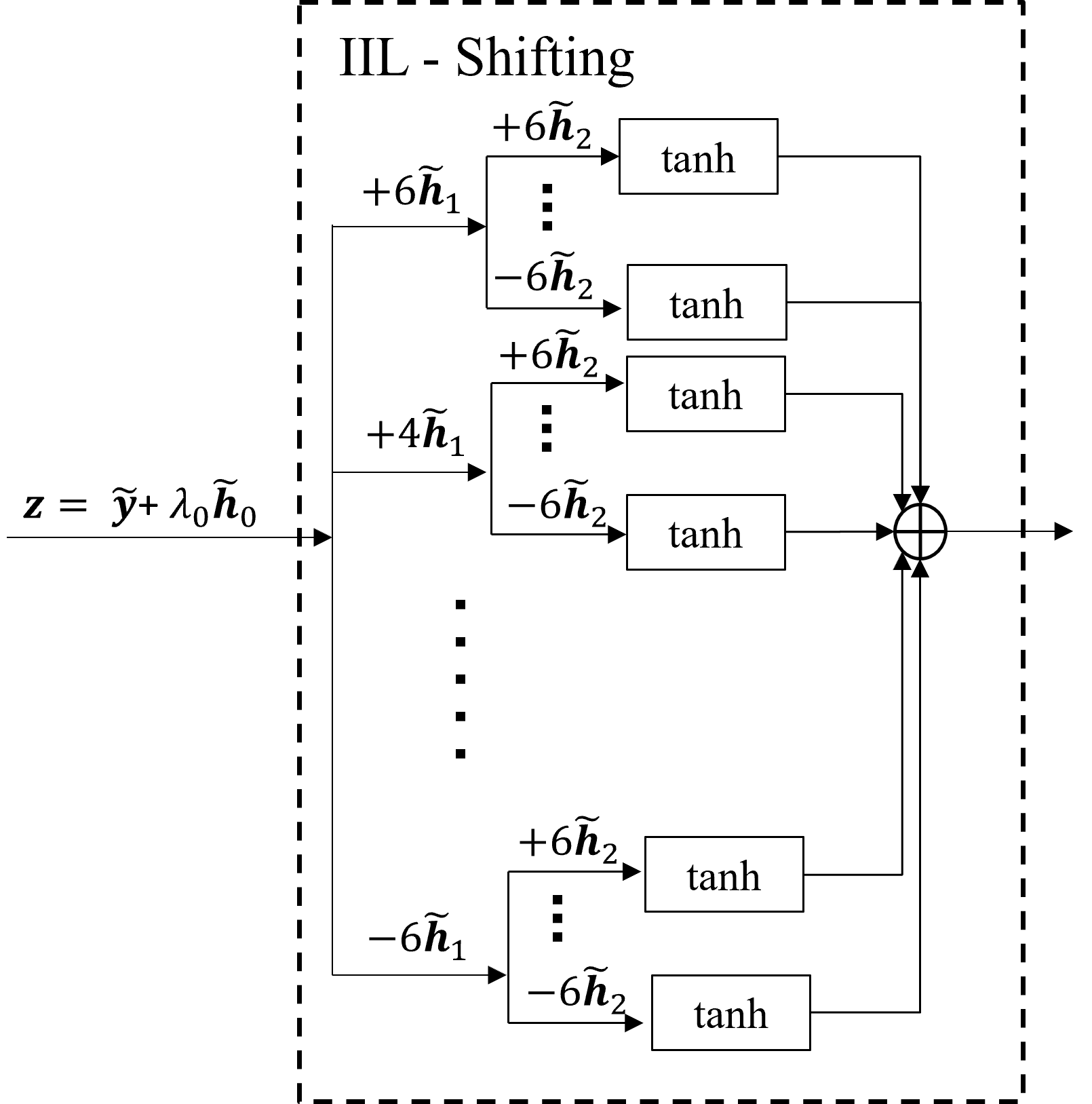}
    \caption{IIL by the shifting solution, three data streams}
    \label{fig:IIL-shift-3}
\end{figure}

Now we design IIL based on this concept, let's start with a simple case where $N_t = 2$, without loss of generality assuming $\tilde{\bm{h}}_0$ is desired channel, then $\tilde{\bm{h}}_1$ is the interference channel. By assigning $g(\bm{z})=\tanh(\bm{z})$, and $N = 2\tilde{\bm{h}}_1$ in equation (\ref{eq:shift-period}), we have:
\begin{equation}
    f_0^{\text{IIL}}(\bm{z}) = \sum_{m \in \mathbb{Z}} \tanh(\bm{z} + 2m\tilde{\bm{h}}_1),
\end{equation}
note above summation is over infinite items, in real implementation we can only approximate this function by constraining $m$ in a finite integer set, e.g.,
\begin{equation}
    f_0^{\text{IIL}}(\bm{z}) = \sum_{m=-3}^{3} \tanh(\bm{z} + 2m\tilde{\bm{h}}_1),\label{eq:IIL-shifting}
\end{equation}
Fig. \ref{fig:IIL-shift-2} illustrates the IIL structure for this two data streams case.
When $N_t > 2$, the IIL design can be generalized as:
\begin{equation}
    f_0^{\text{IIL}}(\bm{z}) = \sum_{m_1} \sum_{m_2} \cdots \sum_{m_{N_t-1}}\tanh(\bm{z} + 2m_1\tilde{\bm{h}}_1 + 2m_2\tilde{\bm{h}}_2 + \cdots + 2m_{N_t-1}\tilde{\bm{h}}_{N_t-1}), \label{eq:shift-period-nt}
\end{equation}
Fig. \ref{fig:IIL-shift-3} illustrates the $N_t=3$ case. It can be seen this solution works well when $N_t$ is small, but as $N_t$ becomes larger, because the number of items to be summed in equation (\ref{eq:shift-period-nt}) increases exponentially with $N_t$, eventually it will lead to an unacceptable computational cost. Therefore, a solution with computational cost linearly increasing with $N_t$ is more desirable, this motivates the modulo solution.

\begin{figure}[htb]
    \centering
    \includegraphics[width=0.5\columnwidth]{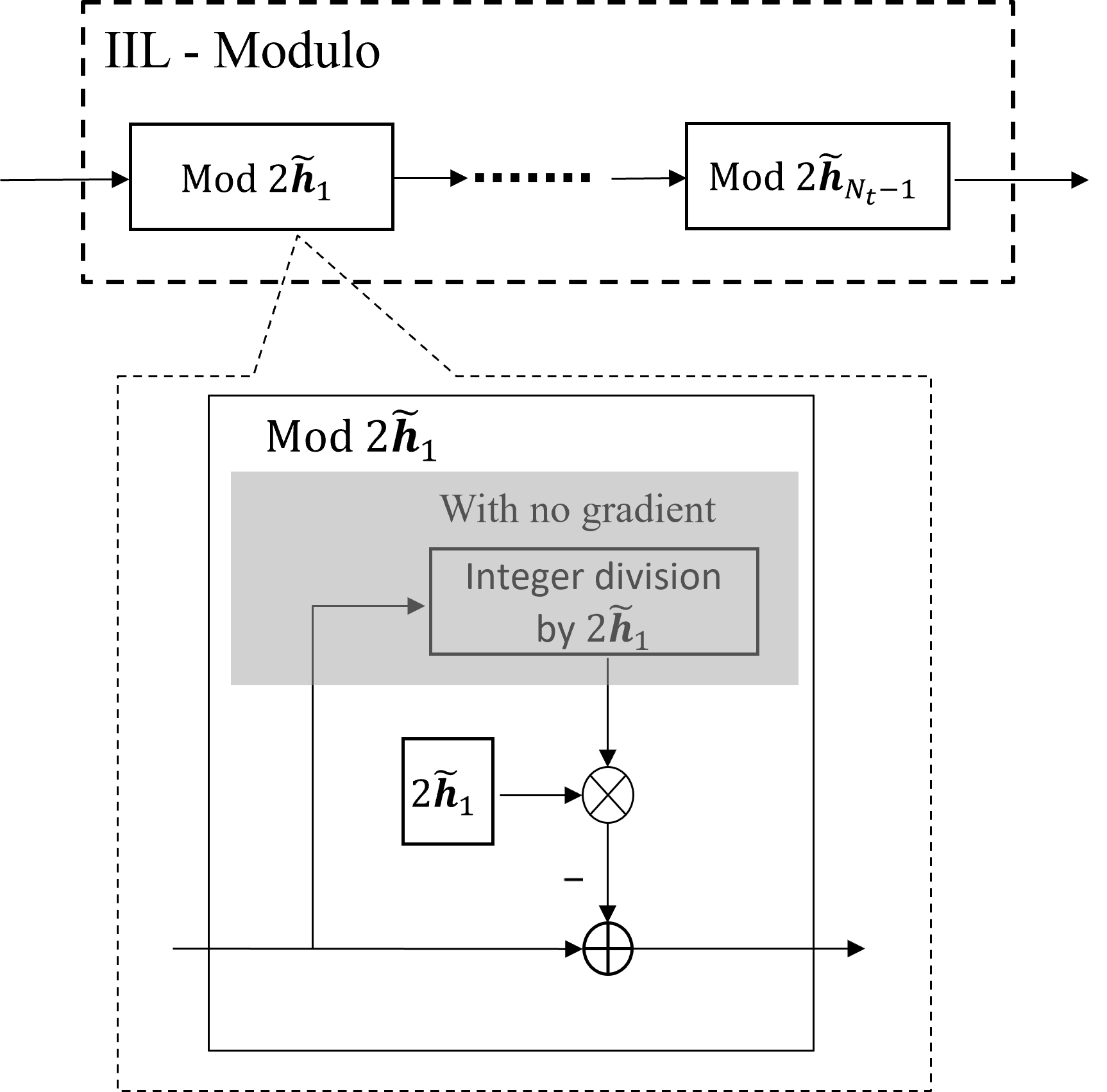}
    \caption{IIL by the modulo solution}
    \label{fig:IIL-modulo}
\end{figure}

\textbf{The modulo solution}: As shown in Fig. \ref{fig:IIL-modulo}, this solution performs mod operation on the input signal with respect to each interference channel sequentially, which can be expressed as:
\begin{equation}
     f_0^{\text{IIL}}(\bm{z}) =  (((\bm{z} \ \text{Mod} \ 2\tilde{\bm{h}}_1) \ \text{Mod} \ 2\tilde{\bm{h}}_2) \ \cdots \ \text{Mod} \ 2\tilde{\bm{h}}_{N_t-1}).
\end{equation}
The benefit of this sequential operation are twofold: first, it makes the computational cost only increasing linearly with $N_t$; second, the sequence order can be easily changed, providing one more knob for performance improvement, e.g., from experiment we found the sequence determined by the descending order of the interference strength gives the best result.

One practical issue when directly using mod function in NN is that the derivative of mod function is not defined, so the training loss can't back propagate into the interference channel coefficients. To solve this issue, we implement mod function with one integer divide, one multiplication, and one subtraction, as shown in Fig. \ref{fig:IIL-modulo}. To be specific,
\begin{align}
    \bm{z} \ \text{Mod} \ 2\tilde{\bm{h}}_j &= \bm{z} - 2\tilde{\bm{h}}_j \cdot \alpha, \label{eq:modulo} \\
    \text{where} \ \alpha &= \lfloor \bm{z} \slash 2\tilde{\bm{h}}_j \rfloor. \label{eq:modulo-alpha}
\end{align}
In this way the training loss can back propagates into the $\tilde{\bm{h}}_j$ in equation (\ref{eq:modulo}). Note the training loss doesn't propagate into $\alpha$, so we don't need to worry about the derivative of the integer divide function (\ref{eq:modulo-alpha}).

\begin{figure}[htb]
    \centering
    \includegraphics[width=0.5\columnwidth]{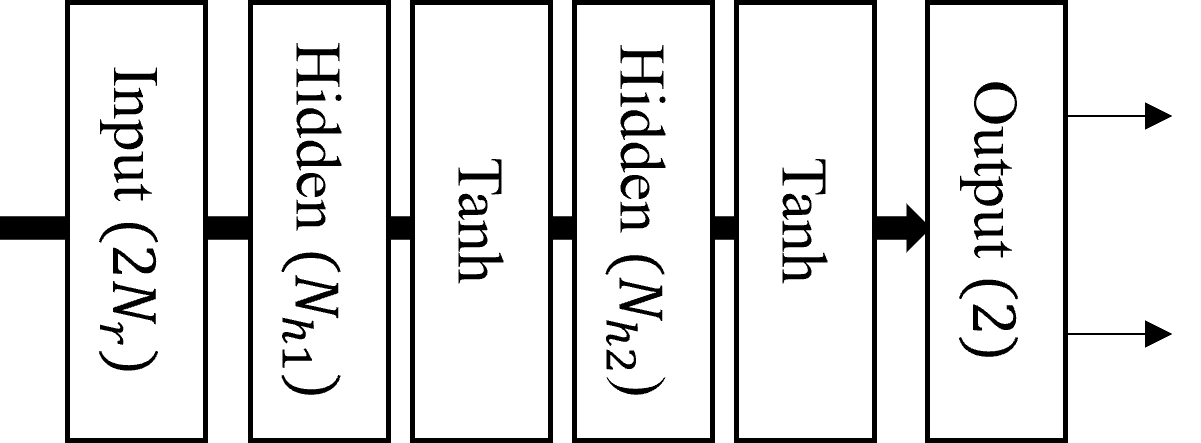}
    \caption{The binary classifier MLP structure}
    \label{fig:mlp}
\end{figure}

\paragraph{Binary classifier}
The binary classifier is implemented as a MLP. Specifically, it has an input layer with size $2N_r$; following are two hidden layers with $N_{h1}$ and $N_{h2}$ neurons respectively, both hidden layers adopt hyperbolic tangent as activation function; finally the output layer generates two values representing the possibilities of been positive and negative respectively. Fig.\ref{fig:mlp} illustrates the MLP structure.

\subsection{Training Procedure}
At each subcarrier, and for each data stream, $N_p$ pilot symbols are utilized to prepare training data. To be specific, for each pilot symbol $(\tilde{x}_i^{p}, \bm{\tilde{y}}^p)$ (where $\tilde{x}_i^{p}$ is transmitted pilot symbol, $\bm{\tilde{y}}^p$ is the received one), two binary training samples are generated, one positive and one negative. The label-input tuple can be expressed as:

\begin{align}
    \{+1, \ \bm{\tilde{y}}^p + (-\tilde{x}_i^p +1)\bm{\tilde{h}}_i \}, \nonumber\\
    \{-1, \ \bm{\tilde{y}}^p + (-\tilde{x}_i^p -1)\bm{\tilde{h}}_i \}. \label{eq:structnet-train}
\end{align}
We can see through shifting the received signal, the transmitted symbol is moved to the positive position ($+1$), and negative position ($-1$) respectively. After passing training input through StructNet-CE, the cross entropy loss is calculated between output and training label. Then the channel coefficients in channel layer and IIL, as well as the binary classifier weights are updated through back propagation. Note the network adopts an alternative learning strategy. When training the binary classifier, the channel weights remain fixed. When updating the channel layer and IIL, the binary classifier weights are unchanged. After training, the channel weights $\tilde{\bm{h}}_i$ and $\tilde{\bm{h}}_j$ are read out and concatenated to form the estimated channel $\hat{\bm{H}}(c)$. The channel estimation procedure is summarized in Algorithm \ref{alg:1}.

\begin{algorithm}
    \caption{Channel Estimation Procedure of StructNet-CE}\label{alg:1}
    \begin{algorithmic}[1]
    \For{Each OFDM subframe}
    \For{Each subcarrier $c$}
    \State Obtain LS channel estimation $\hat{\bm{H}}_{LS}(c)$ by equation (\ref{eq:LS})
    \For{Each data stream $i$}
    \State \multiline{ Initialize the channel layer weights $\tilde{\bm{h}}_i$ and IIL weights $\tilde{\bm{h}}_j$ by utilizing $\hat{\bm{H}}_{LS}(c)$ and equation (\ref{eq:h_tilde})}
    \State Initialize the binary classifier weights following normal distribution
    \State \multiline{Utilizing the transmitted and received pilot symbol pairs $(\tilde{x}_i^{p}, \bm{\tilde{y}}^p)$ to create binary training samples following equation (\ref{eq:rcstruct-test})}
    \For{N training epochs}
    \State Train the binary classifier with symbol detection cross-entropy loss
    \State Train the channel layer and IIL with symbol detection cross-entropy loss
    \EndFor
    \EndFor
    \EndFor
    \State Gather all $\tilde{\bm{h}}_i$s and  $\tilde{\bm{h}}_j$s to form the StructNet-CE estimated channel $\hat{\bm{H}}_{\text{StructNet-CE}}$
    \EndFor
    \end{algorithmic}
\end{algorithm}

\subsection{Complexity}
StructNet-CE is trained through gradient decent, so the computational complexity is proportional to the number of NN parameters. As shown before in Fig. \ref{fig:structnet}, StructNet-CE is comprised of three parts: channel layer, IIL, and classifier. Assuming training with $N_{ep}$ epochs, for channel layer, the complexity is $\mathcal{O}(N_{ep}N_r)$. For IIL, there are two implementation options, for the shifting option, it is $\mathcal{O}(N_{ep}M^{N_t}N_r)$, where $M$ is the cardinality of the finite integer set that $m$ belongs to (\ref{eq:IIL-shifting}). While for the modulo option, the complexity is $\mathcal{O}(N_{ep}MN_tN_r)$. Regarding he binary classifier, it is implemented as a MLP (Fig. \ref{fig:mlp}), so the complexity is $\mathcal{O}(N_{ep}(2N_rN_{h1}+N_{h1}N_{h2}+2N_{h2}))$. All aforementioned complexity is for processing the $i$th transmitter antenna, and in total there are $N_t$ transmitter antennas. Table \ref{table:training-complexity} summarizes StructNet-CE training complexity.

\begin{table*}[htbp]
\centering
\caption{Training complexity}
\resizebox{0.8\columnwidth}{!}{%
\begin{tabular}{l|c} 
\toprule
\textbf{Algorithm}    & \textbf{Complexity per OFDM subframe} \\
\hline
StructNet-CE (Shifting IIL) &$\mathcal{O}\big(N_tN_{ep}(N_r + M^{N_t}N_r + 2N_rN_{h1}+N_{h1}N_{h2}+2N_{h2})\big)$\\
\hline
StructNet-CE (Modulo IIL) &$\mathcal{O}\big(N_tN_{ep}(N_r + MN_tN_r + 2N_rN_{h1}+N_{h1}N_{h2}+2N_{h2})\big)$\\
\bottomrule
\end{tabular}%
}
\label{table:training-complexity}
\end{table*}

\section{Numerical Experiments}\label{sec:experiment}
\subsection{Complexity of Different IIL Implementations} \label{sec:IIL-time-complexity}
\begin{figure}[htb]
    \centering
    \includegraphics[width=0.7\columnwidth]{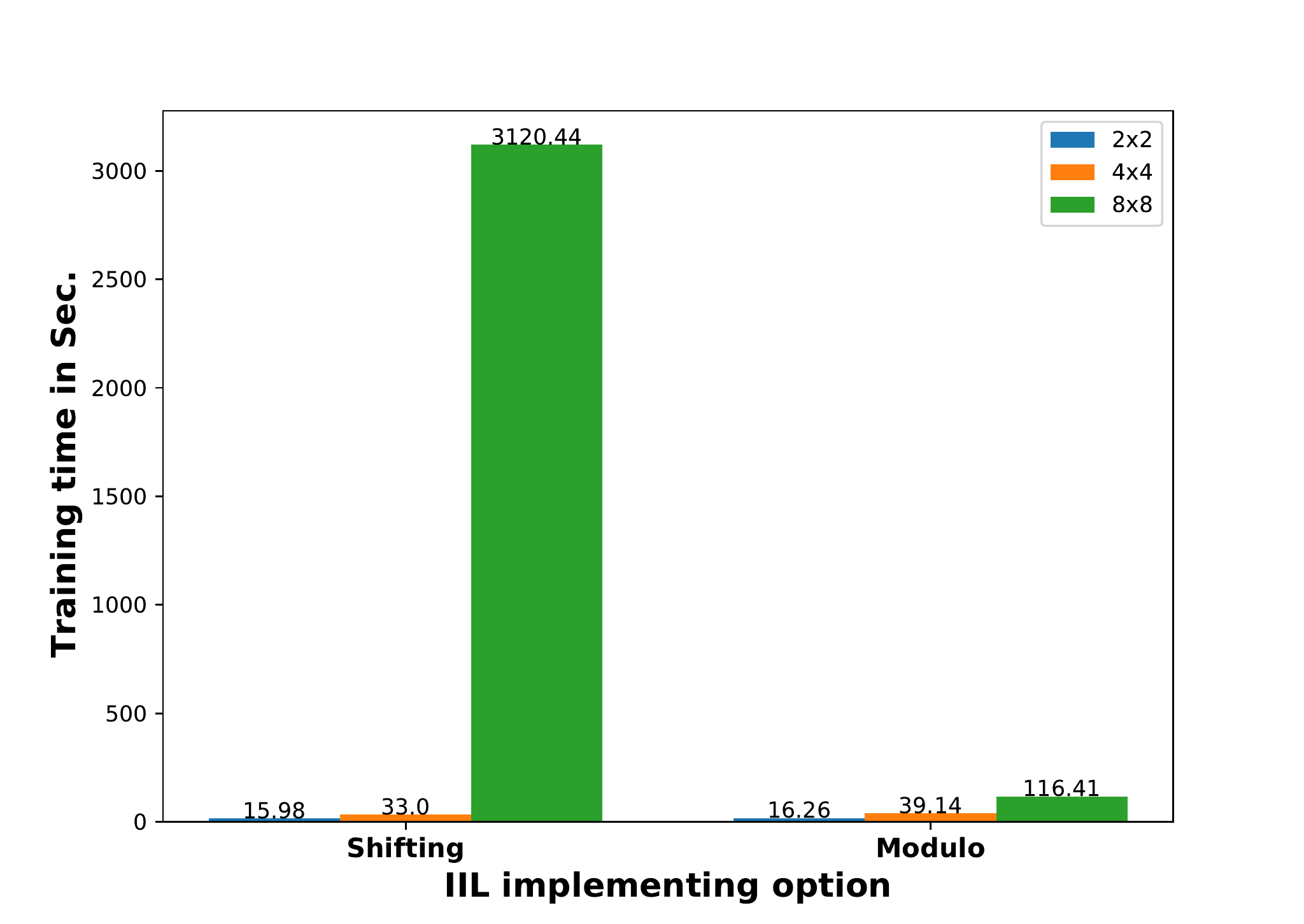}
    \caption{Training time cost (Sec.) of StructNet-CE with different IIL implementing options}
    \label{fig:IIL-time}
\end{figure}

As discussed earlier there are two solutions for IIL implementation, the shifting one, and the modulo one. Theoretically, the computational complexity of the shifting solution increases exponentially with $N_t$, while the modulo one increases linearly. In this section we use a toy experiment to verify this empirically. The toy experiment settings are: the number of subcarriers $N_c = 1$; the number of pilot symbols $N_p=500$, and the number of data symbols $N_d = 3000$; 16QAM is utilized for modulation; the MIMO size varies from $2\times 2$ to $8 \times 8$; training epoch is set to 500; For the shifting solution, we constrain $-3\leq m\leq3$ as in equation (\ref{eq:IIL-shifting}). The experiment was conducted on a desktop PC with Intel Core i5-7400 CPU @ 3.00GHz and 12GB RAM. The training time cost is shown in Fig. \ref{fig:IIL-time}. It can be seen when MIMO size is small, e.g, $2\times 2$ or $4\times 4$, the training time costs for both implementations are almost the same. When MIMO size increased to $8\times 8$, the training time of shifting solution becomes much higher ($26\times$ higher) than the modulo one. And when we increase MIMO size to $16\times 16$, the shifting implementation starts to cause `out of memory' issue on PC. Note the data symbols in this experiment is used to validate the symbol detection performance, make sure the StructNet-CE is working properly. For channel estimation, the data symbols are not needed, that's why we only show training time, omit the testing one. In the rest experiments we choose the modulo solution for IIL implementation due to its low complexity.


\subsection{Channel Estimation Performance}
\begin{figure}[htb]
    \centering
    \includegraphics[width=\columnwidth]{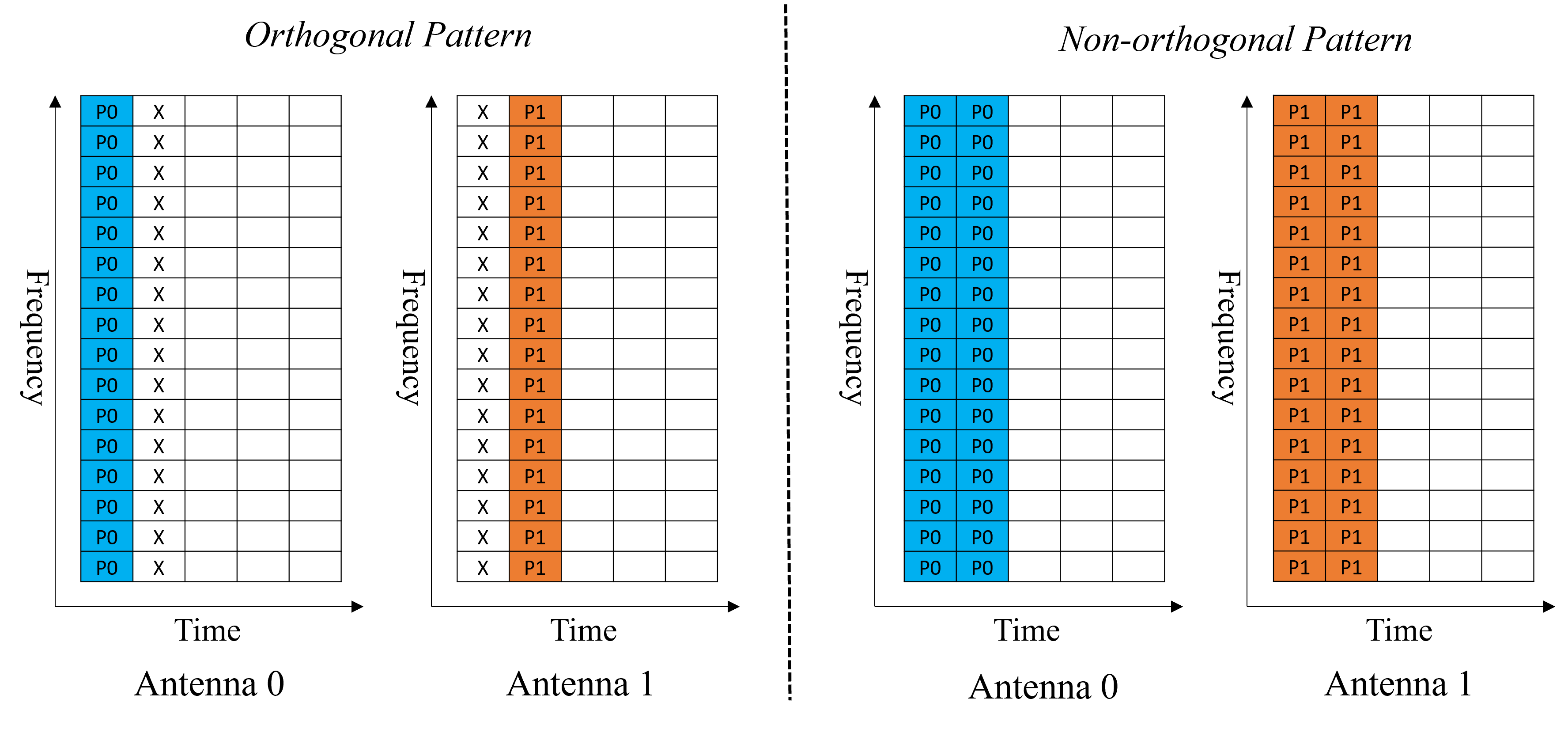}
    \caption{Orthogonal and non-orthogonal pilot pattern}
    \label{fig:pilot}
\end{figure}

In this section we compare channel estimation performance of different methods in terms of MSE and bit error rate (BER). Where MSE is defined as:
\begin{equation}
    \text{MSE} = \frac{\sum_{c=0}^{N_c-1} \|\bm{H}(c) - \hat{\bm{H}}(c)\|_2^2}{N_tN_rN_c}.
\end{equation}
And BERs of different channel estimation methods are calculated based on the same symbol detection method defined in equation (\ref{eq:symbol-detection}), with different estimated channels as input. Two types of pilot pattern are tested in the experiment. As shown in Fig. \ref{fig:pilot}, \textit{orthogonal} pattern means the pilot symbols across different antennas are orthogonal to each other, and the orthogonality is achieved through a time-division multiplexing fashion, i.e., when one antenna transmit pilot symbols, all other antennas stay silent; while in \textit{non-orthogonal} pattern all antennas transmit pilots simultaneously. Pilot symbols are randomly generated QAM symbols.

Regarding the MIMO-OFDM system settings, the MIMO size $N_r = N_t = 2$; number of subcarriers $N_c = 1024$; CP length $N_{cp} = 32$; Each OFDM subframe consists of $N_s = 14$ OFDM symbols, within which $N_p=2$ are pilot symbols and the rest $N_d=12$ are data symbols. The channel realizations are generated with QuaDRiGa version 2.4.0~\cite{jaeckel2014quadriga}, following 3GPP non-line of sight (NLOS) urban macrocell (UMa) channel model~\cite{3gpp.38.901} with central frequency 2.5GHz and bandwidth 10MHz. Table \ref{tab:setting} summarizes all experiment settings. 

\begin{table}[htb]
\centering
\caption{Experiment settings}
\label{tab:setting}
\resizebox{0.8\columnwidth}{!}{%
\begin{tabular}{|c|c|}
\hline
\textbf{Item} &\textbf{Value}  \\ \hline
Number of receiver antenna $N_r$ & 2  \\ \hline
Number of transmitter antenna $N_t$ & 2  \\ \hline
Number of OFDM subcarriers $N_c$ & 1024  \\ \hline
CP length $N_{cp}$ & 32  \\ \hline
Number of OFDM symbols per subframe $N_s$ & 14  \\ \hline
Number of pilot symbols per subframe $N_p$ & 2  \\ \hline
Number of data symbols per subframe $N_d$ & 12  \\ \hline
Channel model & 3GPP UMa NLOS \\ \hline
StructNet-CE binary classifier hidden layer size & $N_{h1}=16, \ N_{h2}=32$ \\ \hline
ReEsNet offline training data size & 100,000 (70\% training, 30\% validation) \\ \hline

\end{tabular}%
}
\end{table}

In total we compare six channel estimation methods, three conventional, and three learning-based. Conventional methods are: LS expressed by equation (\ref{eq:LS}); genie aided LMMSE (genie-LMMSE) defined by equation (\ref{eq:LMMSE}), where we assume perfect channel statistics are known as prior information; em-LMMSE illustrated by equation (\ref{eq:em-LMMSE}), which utilizes estimated channel to calculate empirical channel correlation matrix. Learning-based methods are: StructNet-CE introduced in this paper, which is an online method only utilizes pilot symbols within one OFDM subframe for training, its binary classifier is set to have hidden layer size $N_{h1}=16$ and $N_{h2}=32$; SD-RL \cite{oh2021channel} is a successive denoising (SD) method utilizes RL to learn the denoise order, which also can be seen as an online method, but the nature of RL would require hundreds of OFDM subframes to learn a good policy; ReEsNet \cite{li2019deep} is our previous work designed for channel estimation with comb pilot pattern, which is modified to work with the pilot pattern in this work. ReEsNet requires offline training, for which we generate 100,000 offline channel realizations with mixed signal to noise ratio (SNR) from 0dB to 15dB with 5dB step size, of which 70\% are used for training and the rest 30\% are for validation.

\begin{figure}[htp]
\centering
\subfloat[\textit{orthogonal} pilot pattern]{%
  \includegraphics[clip,width=0.7\columnwidth]{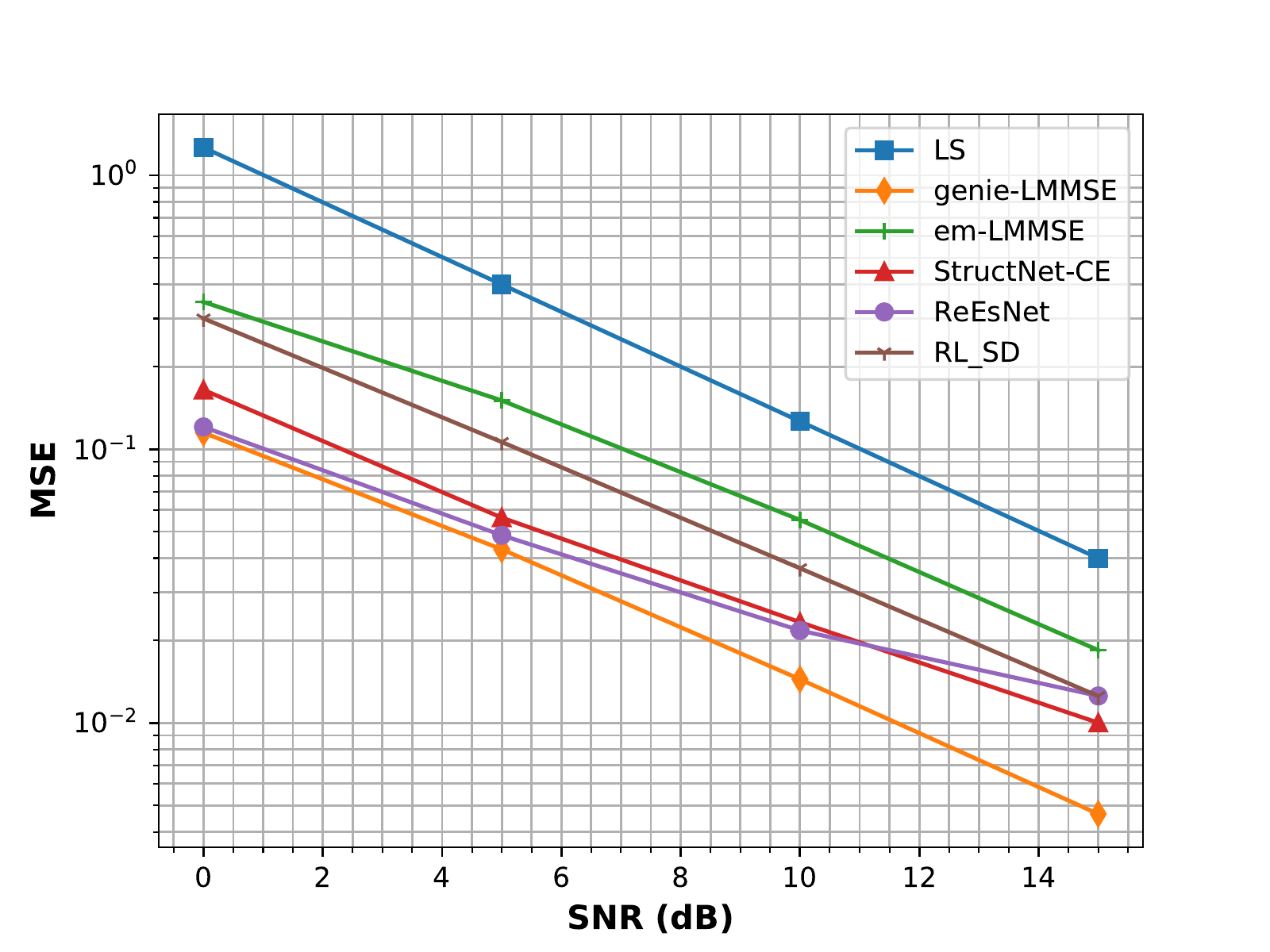}%
}

\subfloat[\textit{non-orthogonal} pilot pattern]{%
  \includegraphics[clip,width=0.7\columnwidth]{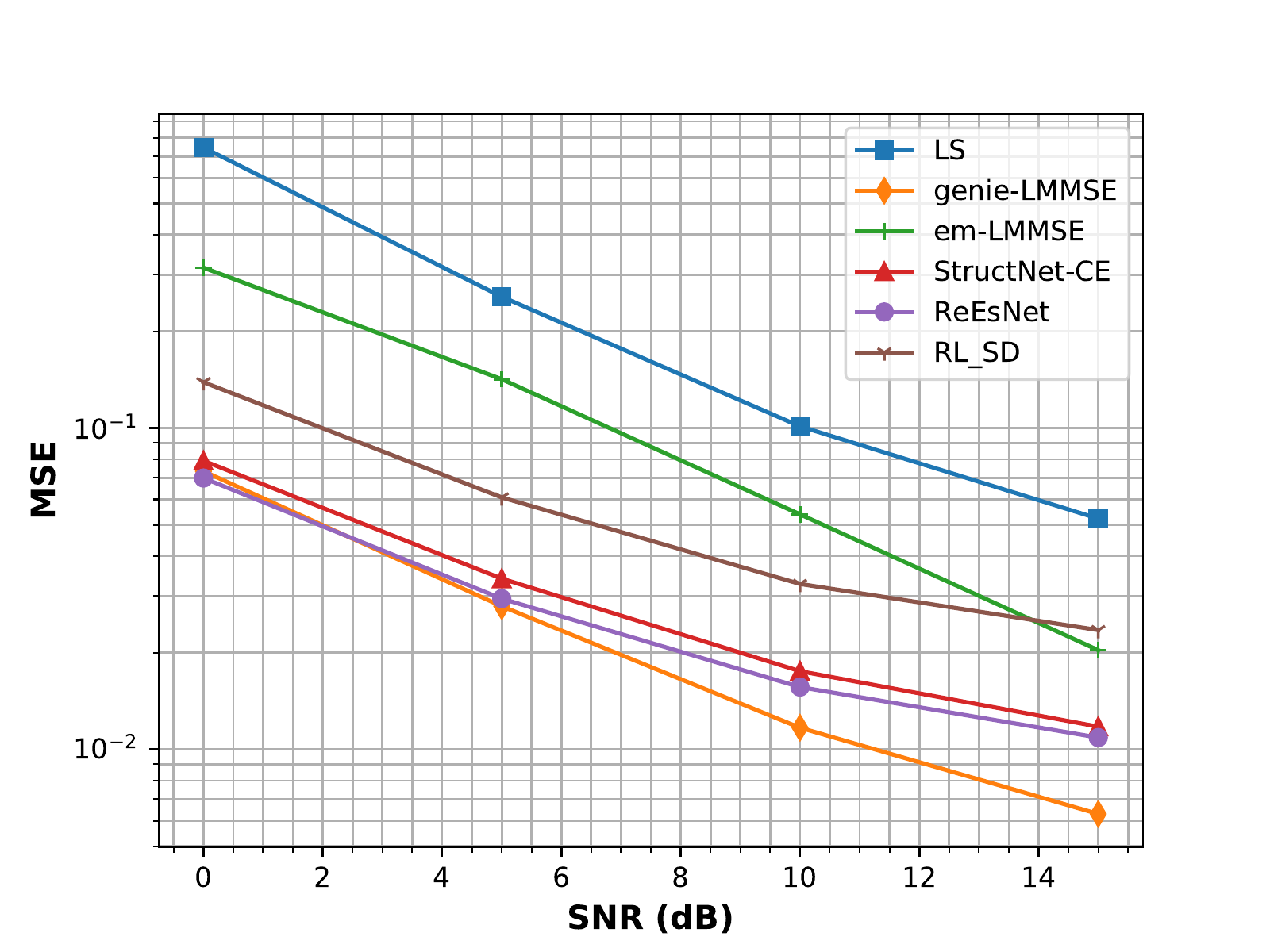}%
}
\caption{MSE of channel estimation methods}
\label{fig:mse}
\end{figure}

\begin{figure}[htp]
\centering
\subfloat[\textit{orthogonal} pilot pattern]{%
  \includegraphics[clip,width=0.7\columnwidth]{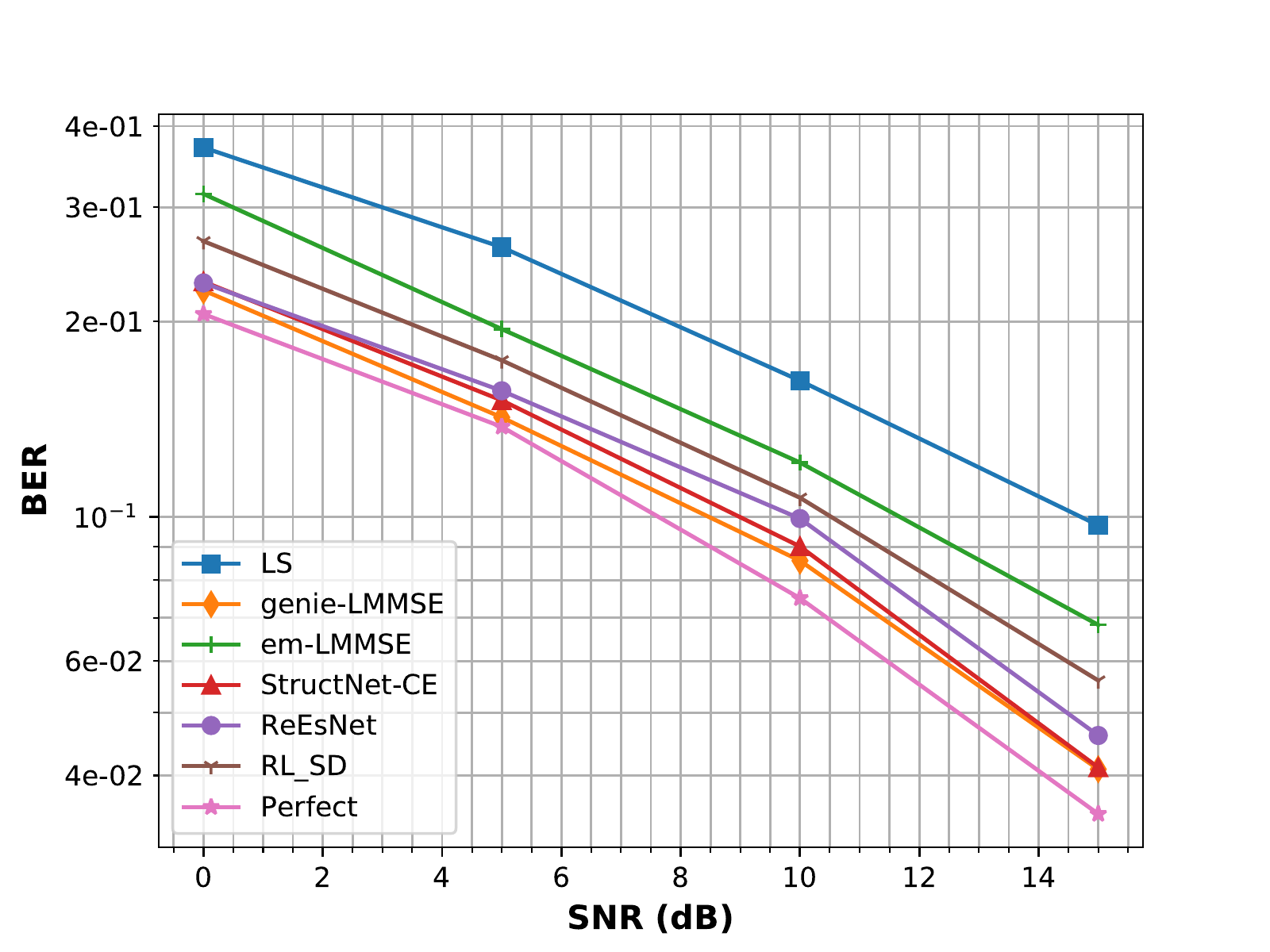}%
}

\subfloat[\textit{non-orthogonal} pilot pattern]{%
  \includegraphics[clip,width=0.7\columnwidth]{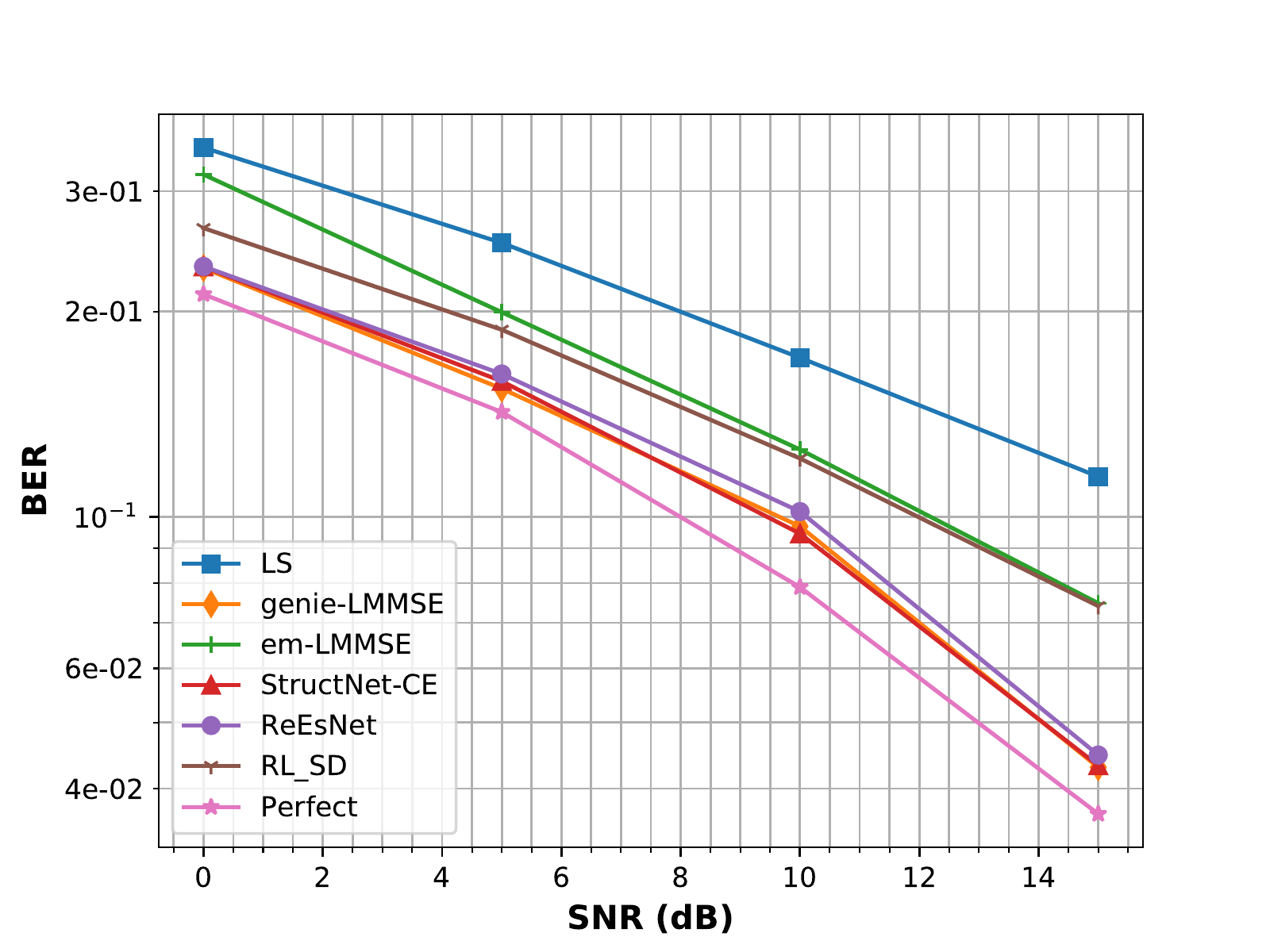}%
}
\caption{BER of channel estimation methods}
\label{fig:ber}
\end{figure}

The MSE of channel estimation methods is shown in Fig. \ref{fig:mse}, with subplot (a) the \textit{orthogonal} pilot pattern, and subplot (b) the \textit{non-orthogonal} pattern. First we can see a general trend that applies to all methods: in the low SNR regime, the \textit{non-orthogonal} pilot pattern has better MSE than the \textit{orthogonal} pilot pattern; while in high SNR regime, the \textit{orthogonal} pilot pattern has better MSE performance. The reason is in low SNR regime noise is the dominant factor, by adopting the \textit{non-orthogonal} pattern each antenna can transmit pilot symbols across all $N_p$ OFDM symbols, then the receiver can essentially average out the noise by utilizing multiple pilot symbols. While in high SNR regime, inter-antenna (inter-stream) interference becomes the dominant factor, which can be totally avoided by adopting the \textit{orthogonal} pilot pattern.
Regarding the relative performance among different methods, LS has the worst MSE. em-LMMSE shows better performance than LS by utilizing estimated channel correlation information. While genie-LMMSE gives the lowest MSE due to the perfect channel statistic knowledge. For leaning-based methods, StructNet-CE outperforms RL-SD with about 2.5dB gain in orthogonal pilot pattern, in non-orthogonal pilot pattern, the gain is about 3.5dB. Although ReEsNet has slightly better performance than StructNet-CE, the offline training making it hardly adopted in any practical wireless communication system.

The BER performance is shown in Fig.\ref{fig:ber}, where subplot (a) is \textit{orthogonal} pilot pattern, and subplot (b) is \textit{non-orthogonal} pattern. On top of those six channel estimation methods mentioned before, we add one more result with perfect channel state information, which is served as performance upper bound. Compare Fig. \ref{fig:ber} with Fig. \ref{fig:mse} we can see in general lower MSE corresponds to better BER, but there are exceptions, for example, StructNet-CE has slightly worse MSE than ReEsNet, but the BER performance is better. Another observation worth mention is that compared with genie-LMMSE, although StructNet-CE has a MSE performance gap about 1dB in low SNR and 3dB in high SNR regime, its BER performance is almost the same as genie-LMMSE across all SNR regimes. After all, the MSE metric only utilizes one value to represent the quality of whole channel estimates, which inevitably losses some information. With the complement of BER metric, we can better understand the performance of channel estimation methods from a different perspective.





\subsection{Empirical Complexity of Channel Estimation Methods}
In this section we show the CPU run time of different channel estimation methods, which empirically reflect their computational complexity. The simulation is conducted on a the same computer as in section \ref{sec:IIL-time-complexity}. The average CPU run time (in second) for processing one OFDM subframe is shown in Table \ref{tab:complexity}. It can be seen LS costs less than one second due to its simplicity. em-LMMSE requires much longer time because it needs to estimate channel statistics, in our implementation 100 subframes is utilized to calculate the channel correlation matrix. Regarding online learning-based methods, StructNet-CE costs around 621 seconds, which is slightly less than em-LMMSE. While RL-SD requires roughly 1385 seconds for processing one subframe, in addition it needs 100 subframes for the RL algorithm to converge, so the total processing time is extremely high.

\begin{table}[htb]
\centering
\caption{CPU run time of symbol detection methods}
\label{tab:complexity}
\resizebox{0.99\columnwidth}{!}{%
\begin{tabular}{cccc}
\toprule
\textbf{Method Type} & \textbf{Method} & \textbf{No. of Subframe to Converge} & \textbf{CPU Run Time (Sec.)} \\ \midrule 
\multirow{2}{*}{\textbf{Conventional}}  &\textbf{LS} &- &0.38  \\
 &\textbf{em-LMMSE} &- &673.06  \\\midrule
\multirow{2}{*}{\textbf{Online Learning}} & \textbf{StructNet-CE} &1 &621.06  \\ 
&\textbf{RL-SD} &100 & 138,548.55 (RL converge) +  1,385.49 (Process one subframe) \\
\bottomrule
\end{tabular}%
}
\end{table}

\section{Conclusion}\label{sec:conclusion}
In this paper, we introduce StructNet-CE to perform channel estimation for MIMO-OFDM systems.
Rather than relying on offline training, our method supports efficient online learning with a limited number of training pilots.
Unlike offline learning-based approaches that rely on the ground truth channel as training label, StructNet-CE is compatible and readily applicable to any practical wireless networks.
Potential applications, such as improving precoding and scheduling algorithms with the StructNet-CE channel estimation, will be addressed in future work.

Our work also demonstrates the effectiveness and importance of combining deep learning techniques with domain knowledge for wireless systems.
By leveraging domain knowledge, the NN structure can be simplified to facilitate efficient real-time online training.
We expect to see more examples of such hybrid approaches in future wireless system design.

\ifCLASSOPTIONcaptionsoff
  \newpage
\fi

\bibliography{IEEEabrv,bib}
\bibliographystyle{IEEEtran}

\end{document}